\documentclass[letter,preprintnumbers,amsmath,amssymb,floatfix]{revtex4}

\usepackage{graphicx}
\usepackage{dcolumn}
\usepackage{bm}

\begin{document}

\title{Low mass dilepton radiation at RHIC}
\author{K. Dusling and I. Zahed}
\affiliation{Department of Physics \& Astronomy, State University of New York, Stony Brook, NY 11794-3800, U.S.A.}
\date{\today}  

\begin{abstract}
In this work we discuss the emission of low mass dilepton radiation 
from a hydrodynamic evolution model of Au-Au collisions and make comparisons 
with recent PHENIX measurements.  
The dilepton emission rates from the hadronic phase are treated at finite temperature and
Baryon density and are completely constrained by broken chiral symmetry in
a density expansion. The rates are expressed in terms of vacuum
correlators which are measured in $e^+e^-$ annihilation, $\tau$ decays
and photo-reactions on nucleons and nuclei.   
We consider two possibilities for the hadronic phase: A chemical equilibrated an off equilibrium hadronic gas.  We find that while chemical off-equilibrium helps explain part of the low mass (0.15 $\leq$ M GeV $\leq$ 0.7) enhancement seen in the data there is still a large discrepancy.  

\end{abstract}

\maketitle  

\section{Introduction}

Dileptons and photons are a particularly interesting observable from heavy ion collisions since electromagnetic probes do not interact with the medium after they are produced and therefore carry information on the early stages of the evolution.  In an ideal world one would hope to have an unambiguous signal coming from the quark gluon plasma in order to verify its existence and study its properties. In any collision, there is also a substantial contribution of dileptons coming from the hadronic phase.  The hadronic emission dominates over the quark-gluon plasma signal and is therefore seen as a large background to one interested in studying the QGP.  However, the hadronic phase does contain interesting physics in itself.  An understanding of the resulting hadronic yields can provide crucial information on modifications to electromagnetic spectral functions due to chiral symmetry restoration.

There is a long history of both experimental and theoretical work in electromagnetic probes, which we don't attempt to summarize here.  Most recently there has been two experiments that have looked at dilepton emission in heavy-ion collisions.  The recent NA60 experiment at the CERN SPS has measured the invariant mass spectrum and the transverse mass spectrum of low-mass dimuon pairs \cite{Arnaldi:2006jq,Damjanovic:2007qm, Damjanovic:2006bd} in In-In collisions.  It was seen that a large excess remained after subtracting contributions from expected hadronic (the cocktail) decays.  The remaining excess was examined by a number of groups \cite{Dusling:2006yv, Dusling:2007kh, vanHees:2006iv, Renk:2006ax, Renk:2006dt, vanHees:2006ng} and was interpreted as a combination of thermal partonic and hadronic contributions with modifications to the spectral function due to finite temperature and Baryon density. 

An experiment at RHIC, performed by the PHENIX collaboration \cite{:2007xw,Toia:2007yr}, measured di-electron pairs in Au-Au collisions.  They find a large excess above the cocktail in the mass region $0.15 \leq $ M (GeV) $ \leq 0.7$ for more central collisions ($N_{part}\gtrsim250$).  For most central collisions ($N_{part}\sim325$) the yield in this mass region is increased by a factor of $\sim 8$ above the cocktail.  

Quickly looking at the two most recent data sets from NA60 and PHENIX one might conclude that the data are inconsistent with each other.  This is not necessarily the case as the two detectors have very complicated acceptances which distort the resulting yields.  Also the space-time evolution and the resulting chemistry is different in In-In and Au-Au collisions.  Therefore, before any conclusions can be drawn, it is necessary to calculate the resulting yields one expects from standard rate equations at RHIC.     

In this work we calculate the resulting di-electron yields from Au-Au collisions at RHIC energies.  In the next section we summarize the rate equations used in both the partonic and hadronic phases.  Then we describe the hydrodynamic evolution model tuned to RHIC collisions.  We consider two scenarios in this work:  A chemically equilibrated hadronic phase as well as a chemical off-equilibrium hadronic phase.  The resulting yields are then compared to the recent PHENIX data. 

\section{Dilepton Rates}

In this section we summarize the dilepton rates used in the analysis.  We should note that these are the same rates used in a previous analysis of the NA60 data \cite{Dusling:2006yv}.  For the partonic contribution above a critical temperature $T_c \approx 180$ MeV  we use the standard leading order q\={q} result for massless quarks:  

\begin{equation}
\label{eq:Bornqqbar}
\frac{dR}{d^4q}=\frac{-\alpha^2}{12\pi^4}\frac{1}
{e^{ q^0/T}-1}\Biglb( N_C 
\sum_{q=u,d,s}e^2_q \Bigrb) \Biglb[ 
1+\frac{2T}{|\Vec{q}|}\ln(\frac{n_+}{n_-}) \Bigrb]
\end{equation}
where $N_C$ is the number of colors, $e_q$ the charge of the quarks, 
and $n_\pm=1/(e^{(q_0\pm|\Vec{q}|)/2T}+1)$.

Below $T_c$ we use the rate equations presented in \cite{Lee:1998um, Steele:1997tv, Steele:1996su} for a hadronic gas at finite temperature and Baryon density which are constrained entirely by broken chiral symmetry:

\begin{equation}
\frac{dR}{d^4q}=\frac{-\alpha^2}{3\pi^3 q^2}\frac{1}
{1+e^{ q^0/T}}(1+\frac{2m^2_l}{q^2})(1-\frac{4m^2_l}{q^2})^{1/2}\Biglb[
-3q^2
\text{Im} {\bf \Pi}_V(q^2)+\frac{1}{f^2_a}
\int{da}{\bf W}^F_1(q,k)+\int{dN}{\bf W}^F_N(q,p)
\Bigrb]
\label{eq:rate}
\end{equation}

where $da$ and $dN$ are the appropriate phase space factors for Mesons and nucleons respectively as outlined in \cite{Dusling:2006yv}.  The term in square brackets is obtained after keeping terms to first order in an expansion of the thermal structure function in both Meson and nucleon density.  Through the use of three-flavor chiral reduction formulas the terms ${\bf W}^F_1(q,k) \text{ and } {\bf W}^F_N(q,p)$ can be expressed in terms of vector and axial spectral densities which are measured in $e^+e^-$ annihilation, $\tau$ decays and photo-reactions on nucleons and nuclei.

\begin{figure}[hbtp]
  \vspace{9pt}
  \centerline{\hbox{ \hspace{0.0in}
\includegraphics[scale=0.7]{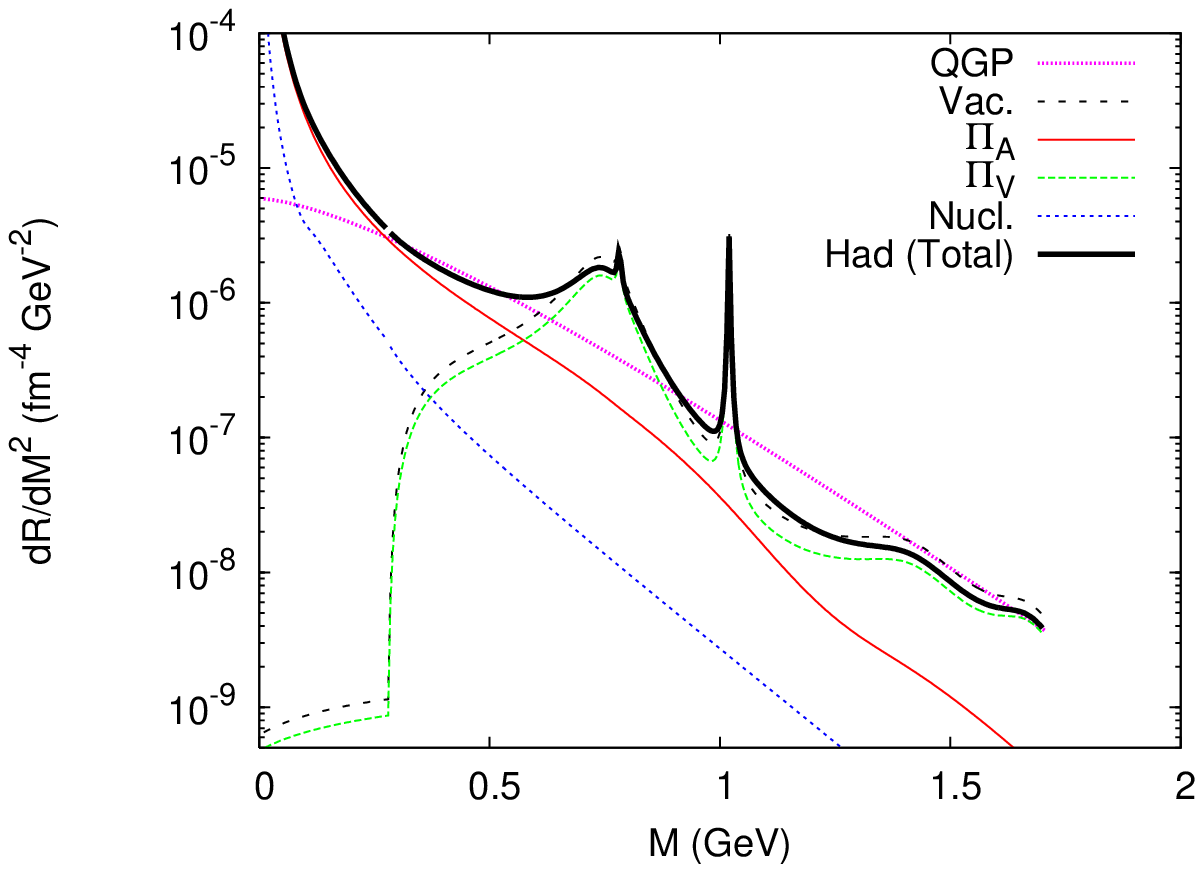}
    \hspace{0.0in}
\includegraphics[scale=0.7]{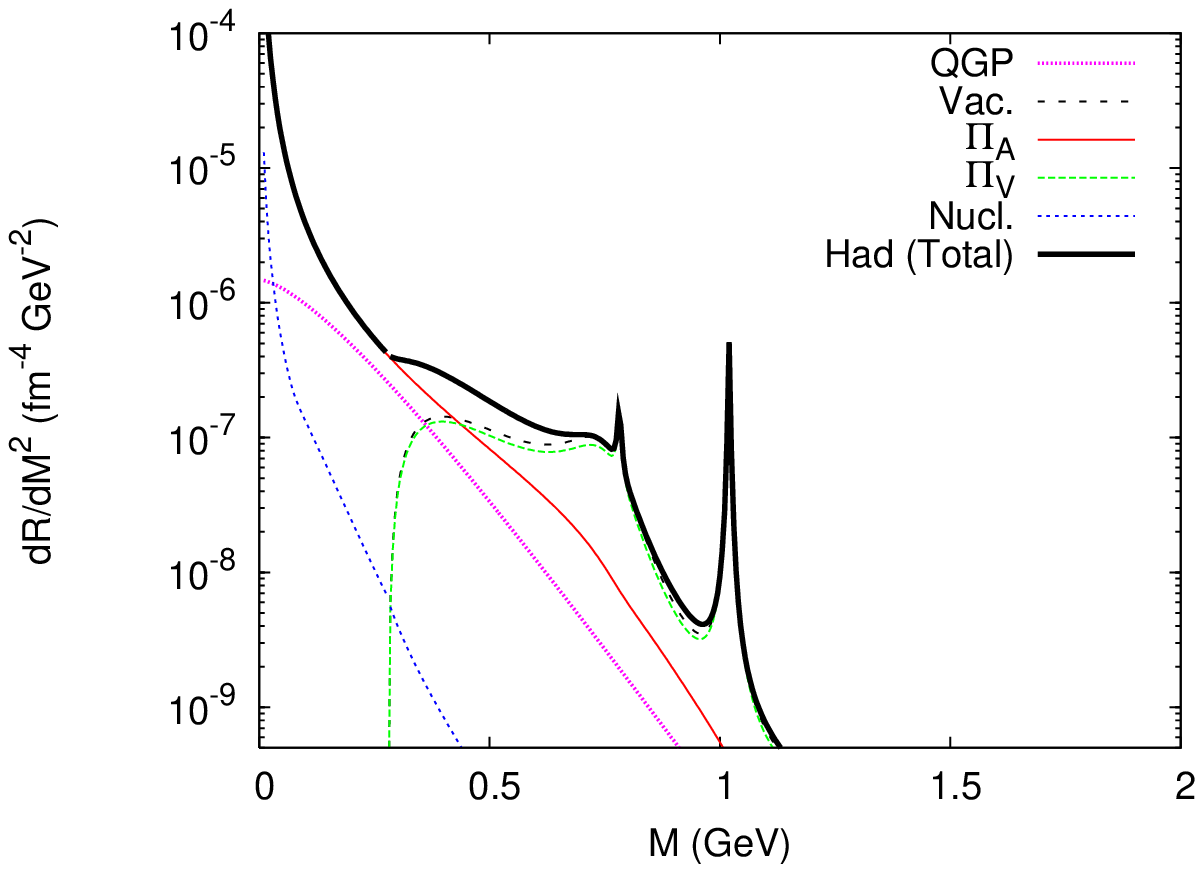}
    }
  }
\caption{
Dilepton rates used in this work.  The QGP is the leading order q\={q} annihilation result.  The curve labeled 'Vac' is the first term in eq.~\ref{eq:rate} and corresponds to the resonance gas contribution.  The curves labeled $\Pi_A$ and $\Pi_V$ correspond respectively to the axial-vector and vector contributions in the density expansion of eq.~\ref{eq:rate}.  
The curve labeled 'Nucl' is the nucleon contribution.   The bold solid curve is the total hadronic rate.  Left: T=180 MeV $\mu_B$=27 MeV Right: T=90 MeV $\mu_B$= 428 MeV, $\mu_\pi$=90 MeV, $\mu_K$=220 MeV.  }
\label{fig:rates}
\end{figure}

In figure~\ref{fig:rates} we show the dilepton rates at two points selected so that one is representative of the beginning of the hadronic evolution and another towards the end.  The left figure shows the rates for T=180 MeV and $\mu_B$=27 MeV which is the conditions at chemical freeezout.  The curve labeled QGP is the leading order contribution from q\={q} annihilation.  The term labeled 'Vac' corresponds to contribution to zeroth order in the density expansion.  The contribution from the term to first order in the density expansion in Mesons is separated into two parts: those from the vector spectral function ($\Pi_V$) and those from the axial ($\Pi_A$).  We should note for clarity that even though we show the QGP contribution in Fig.~\ref{fig:rates} at a temperature of 90 MeV it only exists in our analysis above the critical temperature.  

The main contribution comes from $\Pi_A$ which gives a strong enhancement in the rates below the two-pion threshold.  The mechanism for this can be qualitatively thought of as $X\to\pi + e^+ + e^-$.  It is only qualitative because the physical kinetic processes are mixed by the virial expansion \cite{Lee:1998um}. 

The last piece 'Nucl' is the contribution from finite Baryon densities.  The chemistry at RHIC does not initially support a Baryon rich environment in contrast with fixed target experiments.  Therefore, at chemical freezeout the Baryon density is low and will not significantly modify the rates.  This situation is shown in the left of fig.~\ref{fig:rates}.  Even though the net Baryon number is small at RHIC, the total Baryon number is not.  This observation is taken into account by imposing conservation of both Baryon and anti-Baryon number in the hadro-chemical evolution and will be discussed at length in the following section.  At lower temperatures we find a larger Baryon chemical potential but due to the Boltzmann suppression the rates are still largely unaffected by Nucleons as seen in the right of fig.~\ref{fig:rates}. 
 
\section{Evolution}

In order to see if the rates given in the previous section reproduce the experimentally observed yield, the rates must be convoluted over the full space-time history of the collision region having widely varying temperature, Baryon density, chemical potential and flow velocity.  The evolution of the collision region is treated with a hydrodynamical model based on reference \cite{Teaney:2001av}.  The model is a 2+1 dimensional boost invariant simulation with an ideal gas equation of state ($p=\frac{1}{3}\epsilon$) in the QGP phase and an ideal gas of Meson and Baryon resonances in the hadronic phase. 

The initial condition is set by the entropy in the transverse plane according to the distribution of participants for Au-Au collisions.  One parameter, ($C_s$=entropy per unit spatial rapidity), is adjusted to set the initial temperature and total particle yield. A second parameter ($C_{n_B}$=Baryon number per unit spatial rapidity), is adjusted to fit the net yield of protons.  The parameters used in the hydrodynamic simulation is summarized in table~\ref{tab:param}.

\begin{table}
\begin{tabular}{l|c}
\hline
Parameter & Value \\
\hline
$c^2_{mixed}$   & 0.05c \\
$c^2_{QGP}$     & 0.33c \\
T$_C$   & 180 MeV\\
T$_{f.o.}$      &       120 MeV (90 MeV)\\
$\tau_0$        &       0.6 fm/c\\
$n_B/s$ & 0.004\\
$C_s$   &       15\\
$C_{n_B}$       &       0.06\\
b		&	3.0 fm\\
\hline
\end{tabular}
\caption{\label{tab:param} Parameters used in the hydrodynamic 
simulation of Au-Au collisions at $\sqrt{s}=200$ GeV.  A freezeout temperature of 120 MeV is used for the hadronic phase in chemical equilibrium while T = 90 MeV is used for off-equilibrium.}
\end{table}

For the hadronic phase we consider two different scenarios.  A chemical equilibrated phase ({\em i.e., } vanishing chemical potential for all particle species except Baryons) or a hadronic phase out of chemical equilibrium.  In both cases the relationship between pressure and energy density remain about the same (see Appendix A) and therefore the hydrodynamic solution remains the same.  However, the temperature as a function of energy density changes significantly.  The hadronic phase cools much quicker out of equilibrium.  

As discussed in detail in the Appendix thermal equilibrium is maintained for a much longer time period in the evolution compared to chemical equilibrium in the hadronic phase \cite{Teaney:2002aj}.  This is due to the elastic scattering cross section being much larger then the inelastic cross section.  Elastic collisions do not change the net quantum numbers of particles.  For example the main hadronic reactions, {\em e.g.,} $\pi\pi\to\rho\to\pi\pi, \pi K \to K^* \to \pi K, \pi N \to \Delta\to \pi N$ do not change the net yield of Pions, Kaons and nucleons.  The inelastic collisions which do change the yield of hadronic species take place on time scales longer then the collision time.  In order to take this into account in hydrodynamic simulations which implicitly assume thermal and chemical equilibrium, effective chemical potentials are introduced which keep the particle yields constant throughout the evolution.  Our analysis will therefore consist of two different scenarios.  A chemical equilibrated phase where the chemical potential $\mu$ vanishes for all species except Baryons and a chemical off-equilibrium phase where $n_i/s$ is conserved throughout the evolution.  $n_i$ corresponds to the number density of any hadronic specie listed in the appendix and $s$ is the entropy density. 

There is one further point worth making concerning the role of conserved species.  Rapp \cite{Rapp:2002fc} showed the importance played by conservation of anti-Baryon number on the abundances of other species.  Without anti-Baryon number conservation the Meson chemical potentials tend to remain small. However, when separately imposing anti-Baryon conservation the Meson chemical potentials become quite large.  In our work $\mu_\pi\approx90$ MeV and $\mu_K\approx200$ MeV at $T\approx100$ MeV.  The reason for this is the large amount of entropy stored in B\={B} pairs.  This results in a smaller amount of entropy per Pion therefore requiring a larger $\mu_\pi$ in order to keep $n_\pi/s$ constant.  
  
\begin{figure}
\includegraphics[scale=.8]{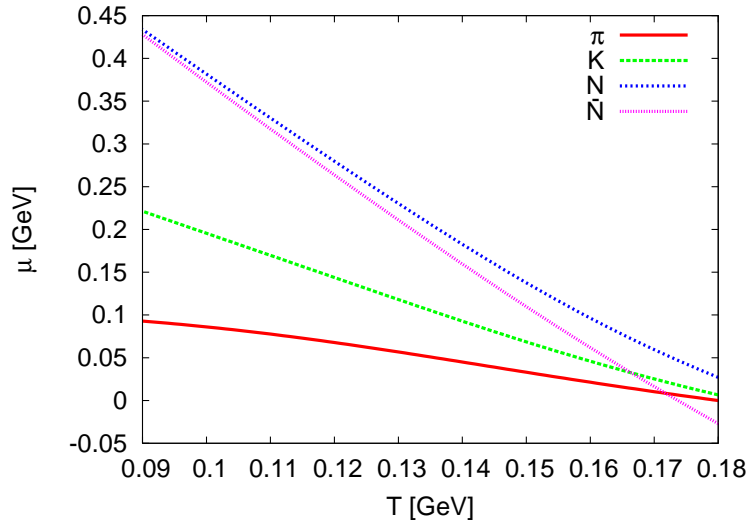}
\caption{Evolution of chemical potentials as a function of temperature.}
\label{fig:eos}
\end{figure}

In figure~\ref{fig:eos} the chemical potential for Pions, Kaons, Nucleons and anti-Nucleons is shown as a function of temperature.  Details of the calculation are discussed in the appendix.  We note that in the chemically equilibrated scenario there is only one chemical potential, $\mu_B$, corresponding to conservation of Baryon number.

\begin{figure}
\includegraphics[scale=.8]{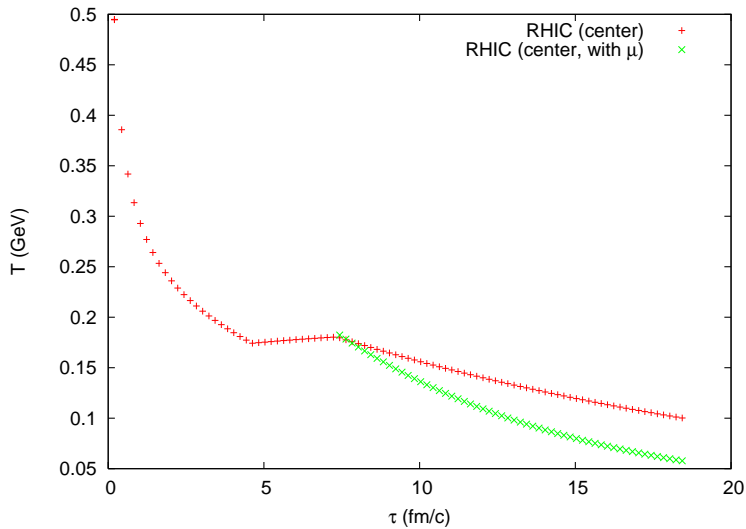}
\caption{Temperature at the center of the hydrodynamic simulation of Au-Au collisions at RHIC as a function of $\tau$ at an impact parameter of b=3 fm.  The curve labeled ('with $\mu$') denotes the trajectory when chemical off-equilibrium is used.  Freezeout occurs in the center at $\tau\sim14$ fm/c.}
\label{fig:Tcen}
\end{figure}

In figure~\ref{fig:Tcen} we show the evolution of temperature at the center of the collision region as a function of proper time.  For the hadronic phase we show two trajectories, one for the chemical equilibrated and the other for the off equilibrium hadronic gas.  We find that the hadronic phase cools quicker in the off-equilibrium (finite $\mu$) scenario.

\begin{figure}
\includegraphics[scale=1.]{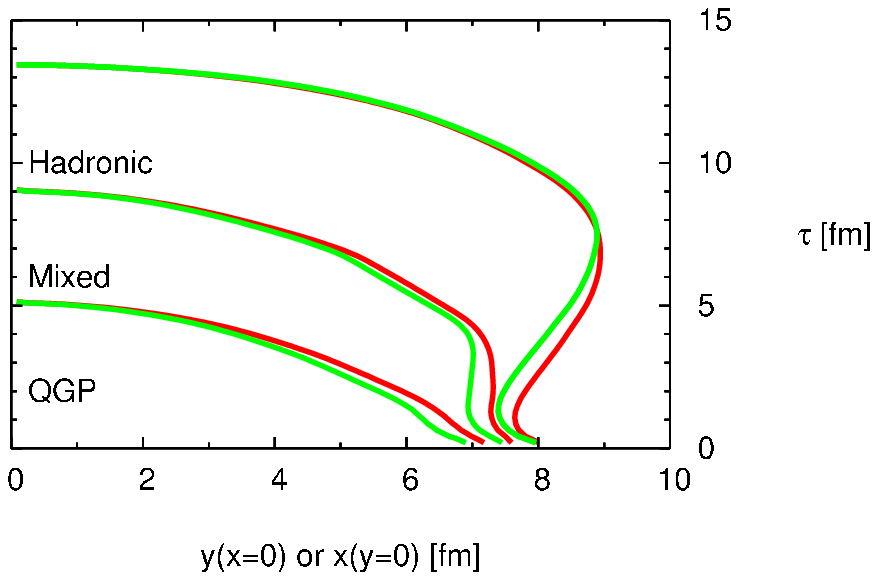}
\caption{The hydrodynamic solution for Semi-Central (b=3 fm) Au-Au collisions at RHIC assuming chemical equilibrium.  The solid lines show contours of constant energy density as a function of y or x at x=0 or y=0 respectively.  The contour values are for $\epsilon_q = 1.8$ GeV/fm$^3$, $\epsilon_h$ = 0.65 GeV/fm$^3$ and for $\epsilon_{f.o.}$ = 0.07 GeV which is corresponds to a kinetic freezeout temperature of T = 130 MeV or T = 90 MeV in and out of equilibrium respectively.  The two curves correspond to slices in the x and y planes of the simulation.}
\label{fig:xy}
\end{figure}

In figure~\ref{fig:xy} the hydrodynamic solution for Semi-Central collisions is shown.  The three contours correspond to the phase transition from the QGP to mixed phase, from the mixed to hadronic phase and the last contour is the space-time location of kinetic freezeout.  

\section{Results \& Discussion}

\begin{figure}
\includegraphics[scale=.8]{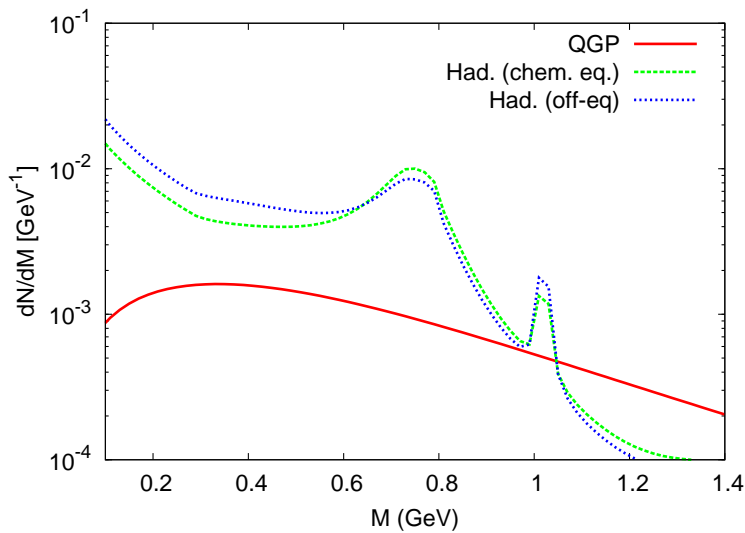}
\caption{Total integrated yield for RHIC showing the partonic contribution and the hadronic contribution with and without chemical freezeout.}
\label{fig:yield}
\end{figure}

We now show the results after evolving the rate equations discussed above through the hydrodynamic evolution.  We show the momentum integrated rates $dN/dM$ for a slice in rapidity $|y|<0.7$ corresponding to the PHENIX detector acceptance in figure~\ref{fig:yield}.  

The partonic contribution does not become dominant until after the $\phi$ mass.  In the low mass region of interest here it is sub-dominant by more than a factor of two.  The overall yields from the two different scenarios used in the hadronic phase are comparable.  This should be expected.  Even though the fireball temperature drops quicker in the off-equilibrium scenario the dilepton rates are enhanced by additional fugacity factors.  Near the $\rho$ peak at M=770 MeV this corresponds to a factor of $z^2$ where $z=e^{\mu_\pi/T}$ which compensates from the reduction due to the Boltzmann factor.  Below the $\rho$ peak the yields are enhanced by a factor of $\approx z^3$.  The extra factor of $z$ comes from the Meson phase space factor in eqn~\ref{eq:rate}.     

This difference in fugacity factors near and below the $\rho$  mass region is what modifies the shape of the hadronic yields when including off-equilibrium effects.  The larger Pion and Kaon densities in the off-equilibrium hadronic phase shuffles strength from the rho peak into the low mass region.  At $M\approx 0.4$ MeV we see about a 50\% enhancement in the yields.  

\begin{figure}
\includegraphics[scale=.8]{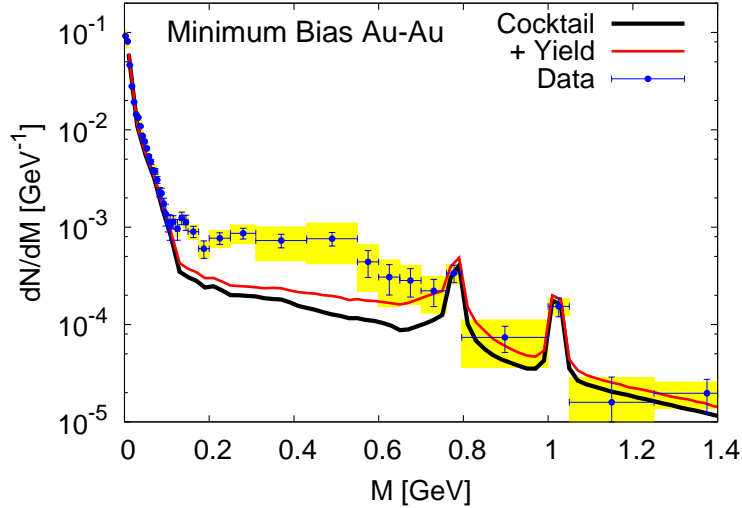}
\caption{Total integrated yield for RHIC showing the sum of partonic contributions and the hadronic contribution with chemical freezeout after a rescaling by the number of participants to fix the overall yield.}
\label{fig:yieldacc}
\end{figure}

In figure \ref{fig:yieldacc} we show the yields from the chemical off-equilibirum scenario after being evolved through the full detector cuts at PHENIX.  The overall yield is determined by rescaling the results by the number of participants corresponding to the minimum Bias data set.  In going from our central collision to min. bias this corresponds to a factor of $\sim N_{part}(min. bias)/N_{part}(central)=109/325$. Ideally we should compare our yields to the 10\% centrality class, which will be done when the data is published.  Even though chemical off-equilibrium helped explain part of the excess di-electrons in the low mass region there is still a large part of the yield that is unexplained.  In an attempt to reconcile this we have tried one further scenario.  A chemically super-saturated phase throughout the entire hadronic lifetime.  A similar analysis was performed in the framework of a Boltzmann type transport model by \cite{Kampfer:1993dk}.  In this model it is assumed that the hadronic phase starts as a supersaturated pion gas with large effective potentials.

\begin{figure}
\includegraphics[scale=.8]{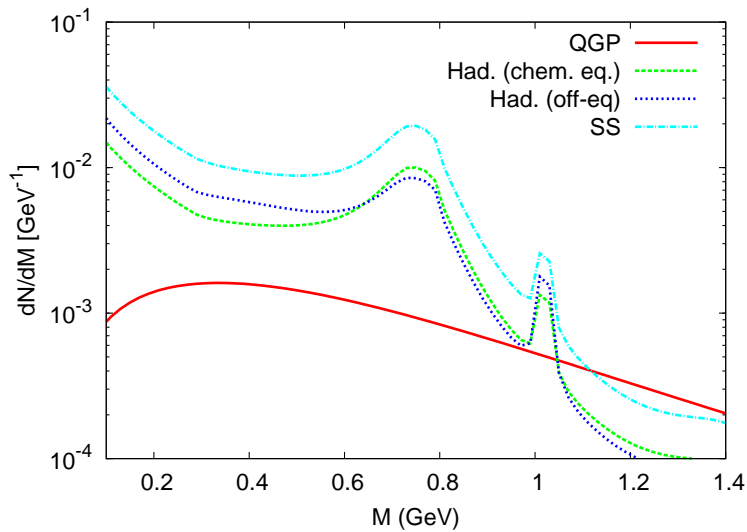}
\label{fig:yieldSS}
\caption{Total integrated yield for RHIC showing the partonic contribution and the hadronic contribution with and without chemical freezeout.  Also shown is the yield using a schematic supersaturated pion gas.}
\end{figure}

We perform this calculation schematically in order to get a qualitative picture of the result.  We use the same evolution model as was used for the chemically equilibrated phase but introduce a pion chemical potential of $\mu_\pi=50$ MeV in the rates which remains constant during the entire phase.  In a more realistic picture one would also need to include this chemical potential in the hydrodynamic equation of state as well.  This would cause the hadronic phase to cool much quicker.  Since we are not doing this we obviously expect to overestimate the yields.  Yet it is still interesting to see how the inclusion of such a chemical potential changes the shape of the spectrum.  This result is shown in fig.~\ref{fig:yieldSS} by the curve labeled 'SS'.  

We should mention that R. Rapp \cite{Rapp:2000pe,Rapp:2002mm} has also performed a similar analysis which included thermal partonic and hadronic contributions.  The hadronic contribution contains medium modifications in the form of self-energy corrections to the $\rho, \omega$ and $\phi$ propagators from resonant interactions with the surrounding Mesons and Baryons as well as corrections from polarization from the pion cloud.  Even though the form of the medium modifications differ in the two calculations the resulting yields are comparable.   

\section{Conclusions}

In conclusion we have calculated the yields of di-electron pairs expected from thermal sources from Au-Au collisions at RHIC.  We have considered a chemically equilibrated and a chemically off-equilibrium hadronic phase.  Even though off-equilibrium effects help explain part of the low mass enhancement seen in the data there is still a large discrepancy.  We have also shown the results from a schematic super-saturated pion gas, which when the correct normalization is taken into account, will probably have only a minimal effect in explaining the excess. 
\\
\\
{\bf Acknowledgments.}
\\
We would like to thank Alberica Toia for running our Yields through the PHENIX acceptance.  KD would like to thank D. Teaney for help implementing chemical freezeout.  This work was partially supported by the US-DOE grants
DE-FG02-88ER40388 and DE-FG03-97ER4014.

\appendix

\section{Equation of State}

In this appendix we discuss some of the details in generating the equation of state used in the above analysis.  The work below is based on previous calculations for SPS done by D. Teaney \cite{Teaney:2002aj}.  The necessity to introduce a separate chemical freezeout temperature arises because of the two different time scales present in heavy ion collisions: $\tau_{th}$ and $\tau_{ch}$.  The first timescale is the thermal equilibration time, $\tau_{th}\sim2$ fm/c at $T\approx160$ MeV and the second is the chemical equilibration time $\tau_{ch}\sim200$ fm/c.  The time scale of a typical heavy-ion collision is on the order of $\tau_{hic}\sim10$ fm/c.  In using a hydrodynamic description it is implicitly assumed that $\tau_{th}\ll \tau_{hic} \ll \tau_{ch}$. 

One therefore has the following picture:  At the critical temperature hadronization occurs and the chemical composition of all species is fixed.  Since the time scale for chemical equilibrium is much longer then the lifetime of the collision the density of the particle species remain fixed (causing the particles to develop chemical potentials as the temperature decreases).  The hadronic fluid remains in kinetic equilibrium due to elastic collisions until a final freezeout temperature when the hydrodynamic simulation terminates and we assume free streaming particles remain in the final state. 

The equation of state during the hadronic phase is constructed by assuming additional conservation laws for the following currents:

\begin{equation}
B, s, I, \bar{s}+s, \pi, K, \eta, \omega, Y, \bar{Y}, \Xi, \bar{\Xi}, \eta^\prime, \Omega, \bar{\Omega}, \phi, \bar{B}
\end{equation}

We note that the conservation of anti-Baryons was not considered in \cite{Teaney:2002aj} but was taken into account due to its importance as pointed out in \cite{Rapp:2002fc}.

The procedure used to generate the equation of state is as follows.  First at $T_c$ the values of $\mu_B, \mu_I$ and $\mu_s$ are fixed so that $s/n_B=250$, $n_I=0$ and $n_s=0$.  Then calculate $n_H/s$ for the remaining 14 densities using their chemical equilibrated ($\mu=0$) values.  Then in small increments the temperature is lowered and the chemical potentials adjusted in order to keep $n_i/s$ constant.  

\begin{figure}[hbtp]
  \vspace{9pt}
  \centerline{\hbox{ \hspace{0.0in}
\includegraphics[scale=0.7]{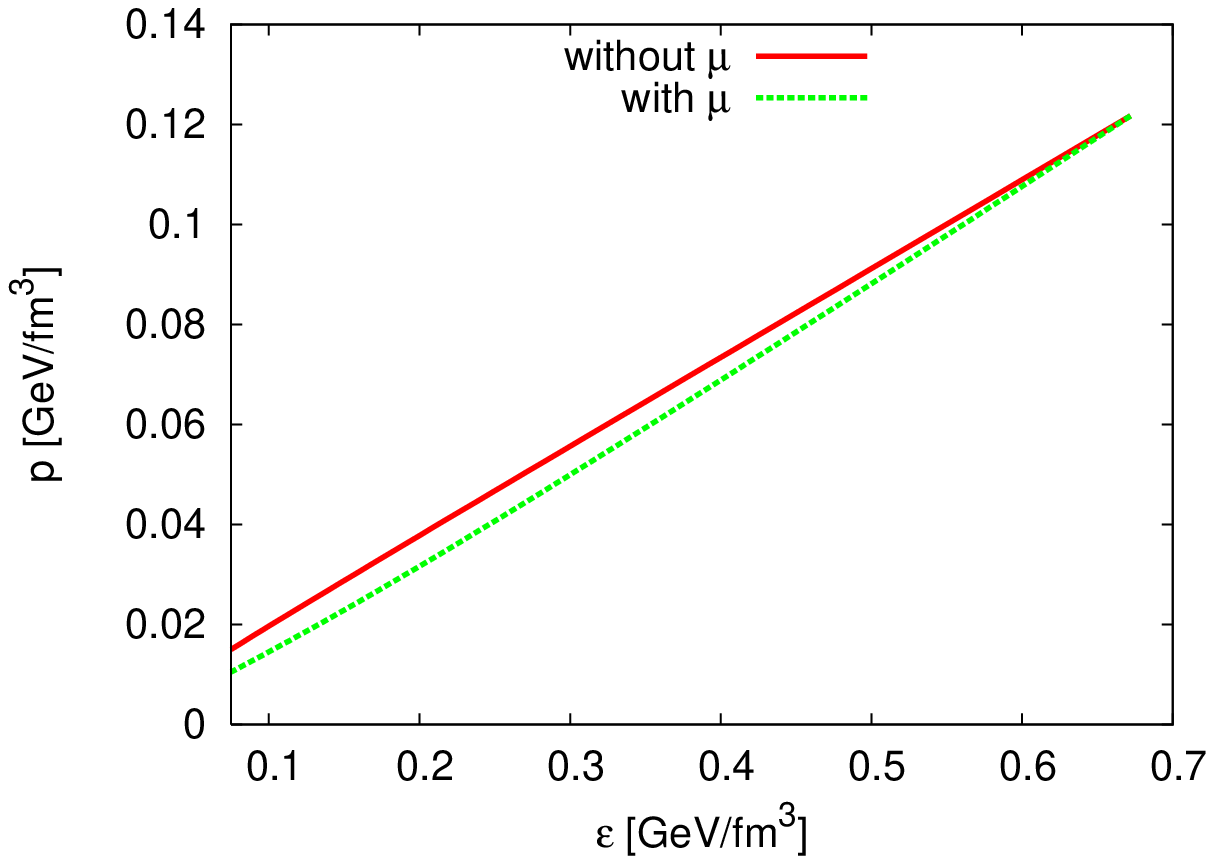}
    \hspace{0.0in}
\includegraphics[scale=0.7]{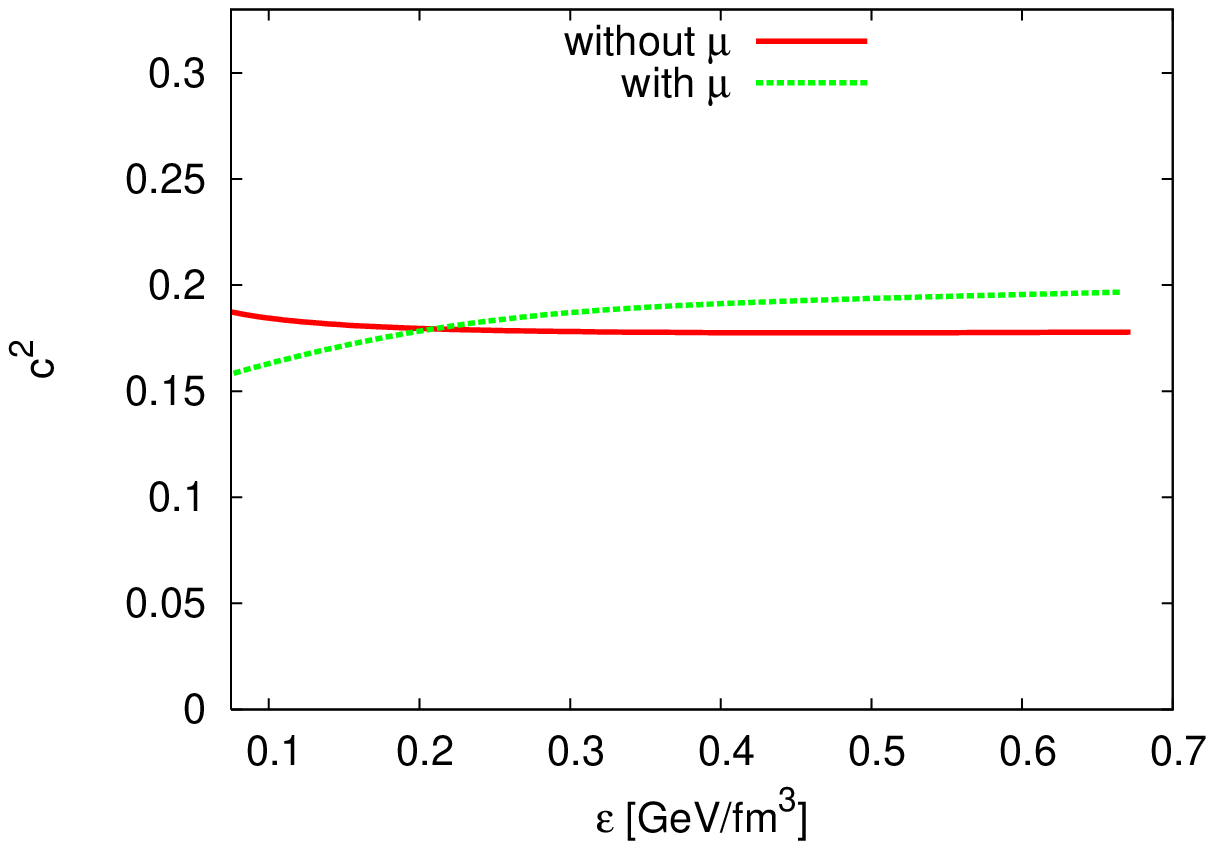}
    }
  }
\caption{The pressure (left) and sound speed squared (right) as a function of energy density with and without chemical freezeout at RHIC ($s/n_B=250$).}
\label{fig:evp}
\end{figure}

We now discuss the results of this procedure.  In figure~\ref{fig:evp} we show the pressure and speed of sound as a function of energy density.  We see that chemical equilibrium does not change the result by much.  We can therefore assume that the hydrodynamic solution does not change and use the same result for both the equilibrated and un-equilibrated phase making sure to use the appropriate temperature when calculating the rates and freezeout.

\begin{figure}[hbtp]
  \vspace{9pt}
  \centerline{\hbox{ \hspace{0.0in}
\includegraphics[scale=0.7]{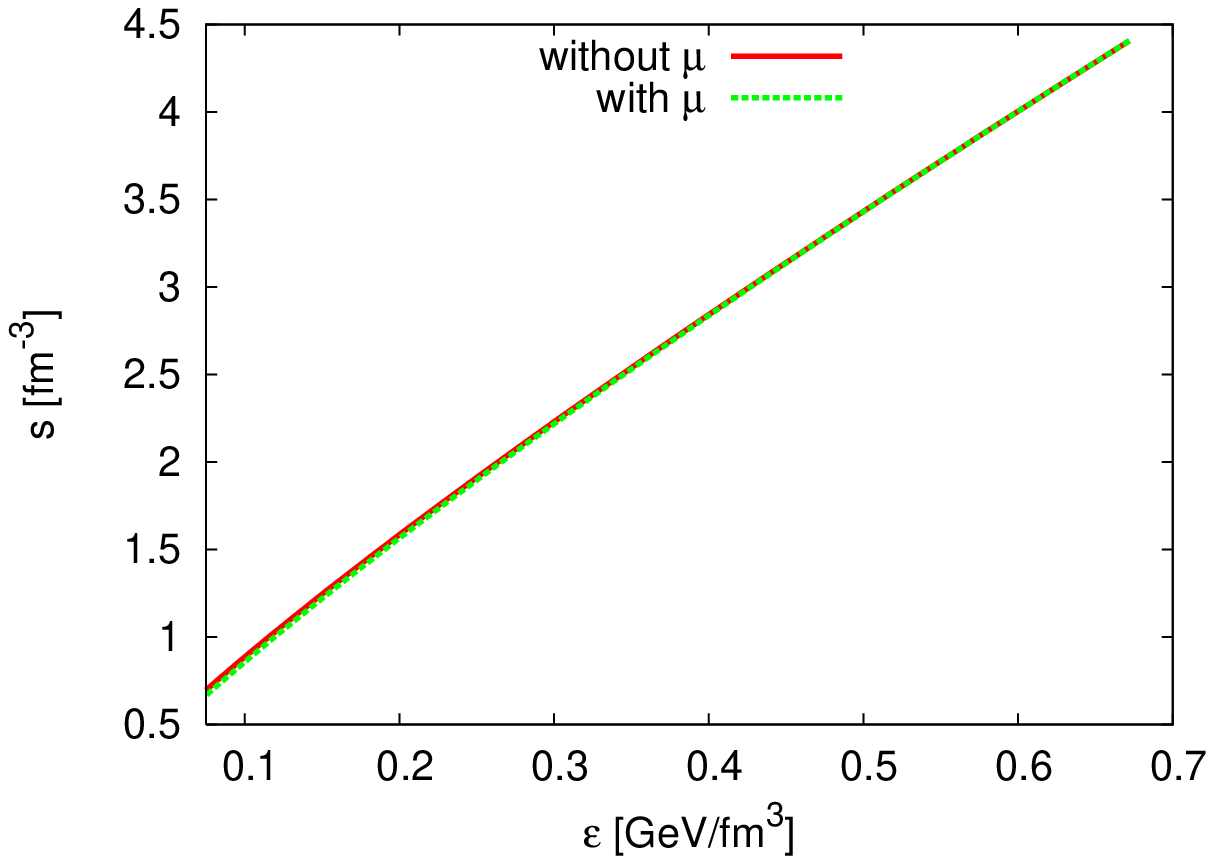}
    \hspace{0.0in}
\includegraphics[scale=0.7]{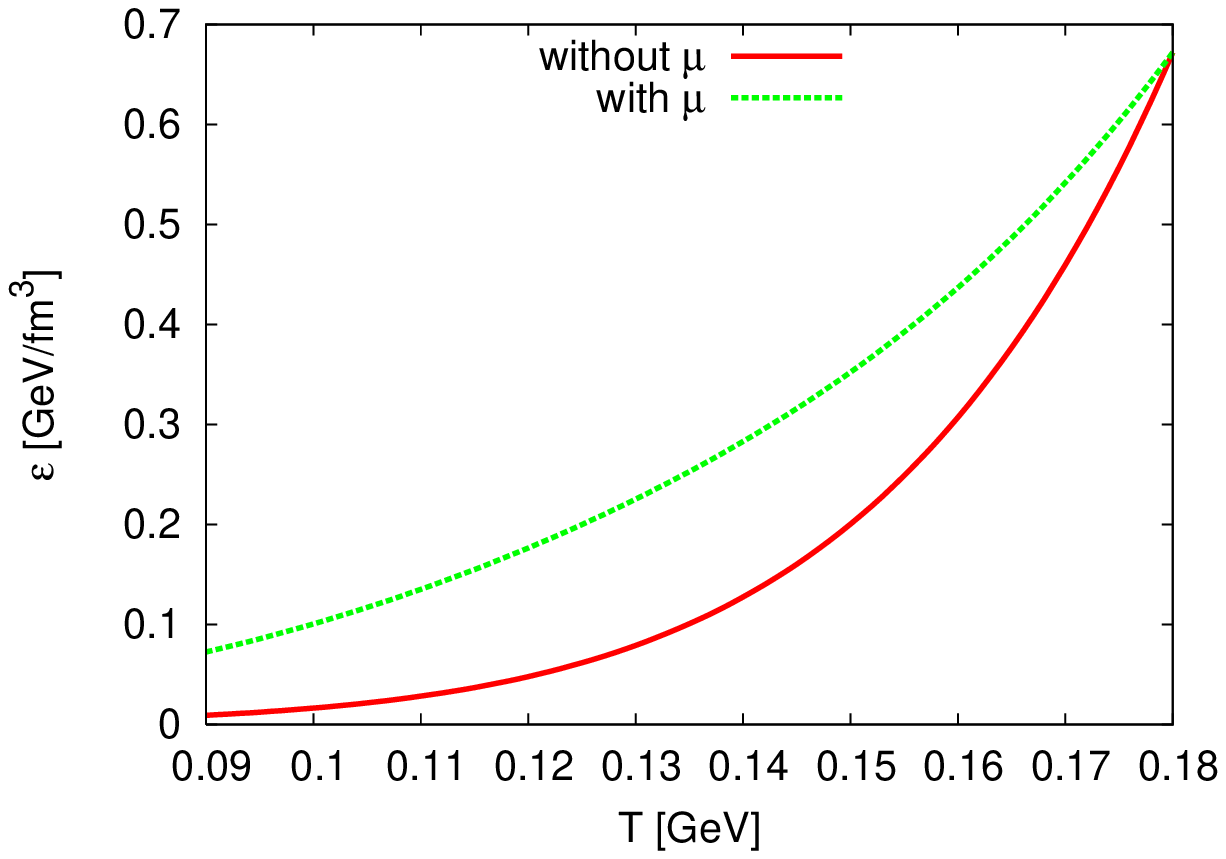}
    }
  }
\caption{Entropy density as a function of energy density (left) and energy density as a function of temperature (right) with and without chemical freezeout at RHIC ($s/n_B=250$).}
\label{fig:evT}
\end{figure}

The temperature as a function of energy density is shown in figure~\ref{fig:evT}.  We see that by introducing chemical freezeout the simulation cools much quicker. This will cause freezeout to occur {\em earlier} in the simulation and lead to shorter lifetimes.  The smaller space-time volume will be compensated for in the yields by the fugacities in the rates.  We also show the entropy density as a function of temperature in order to check our numerics.  This should be the same both in and out of equilibrium, which is indeed the case as shown in figure~\ref{fig:evT}.

The final result for the chemical potentials as a function of temperature is shown in figure~\ref{fig:eos}.  A discussion of the results is kept to the text as it is relevant to the results.

\end{document}